\documentclass[]{spie}  

\makeatletter
\newcommand{\pseudoeqref}[1]{\textup{\tagform@{#1}}}
\makeatother
\usepackage{caption}
\usepackage{subcaption}

\usepackage{amsmath,amsfonts,amssymb}
\usepackage{graphicx}
\usepackage[colorlinks=true, allcolors=blue]{hyperref}

\usepackage{siunitx}
\usepackage{amsmath}
\usepackage{float}
\usepackage{adjustbox}
\usepackage{pgfplots} 
\pgfplotsset{compat=1.3}
\usepackage{url}
\usepackage{todonotes}
\usepackage{gensymb}

\title{Impact of Transceiver Selection on Synchronization Accuracy in White Rabbit Networks}

\author[a]{Michal Špaček}
\author[b]{Josef Vojtěch}
\author[a]{Jaroslav Roztočil}
\affil[a]{Dept. of Measurement, Czech Technical University in Prague}
\affil[b]{CESNET, Prague, Czech Republic}

\authorinfo{Further author information: (Send correspondence to M. Špaček)\\
M. Špaček: E-mail: spacemi6@fel.cvut.cz\\
J. Vojtěch: E-mail: josef.vojtech@cesnet.cz\\
J. Roztočil: E-mail: roztocil@fel.cvut.cz\\\\
\textcopyright\;(2024) Society of Photo‑Optical Instrumentation Engineers (SPIE). One print or electronic copy may be made for personal use only. Systematic reproduction and distribution, duplication of any material in this publication for a fee or for commercial purposes, and modification of the contents of the publication are prohibited.
This is the author-prepared version. The official version is available at: \url{https://doi.org/10.1117/12.3029202}}

\begin{document} 
\maketitle

\begin{abstract}
Achieving optimal synchronization accuracy between two White Rabbit devices hinges on the proper selection of transceivers, which act as electro-optical converters connecting WR devices to the optical network infrastructure. The correct choice of transceivers can significantly improve resilience to changes in the time offset between WR devices due to temperature fluctuations in the connecting optical fiber. To compare the performance of BiDi WDM and DWDM transceivers, an experimental setup was established under laboratory conditions to simulate a real optical network used for distributing precise time and frequency between two remote locations. The optical connection was emulated by integrating a 20 km G.652.D optical fiber into a climatic chamber, which provided variable environmental conditions similar to those experienced in real applications. The study compared BiDi WDM 1310/1550 nm transceivers with DWDM Ch33/Ch34 transceivers. Results showed that DWDM transceivers exhibited nearly thirteen times less sensitivity to temperature-induced changes in the optical connection, leading to a smaller time offset. Therefore, for achieving the highest accuracy in synchronizing WR devices in practical applications, DWDM transceiver technology is essential.
\end{abstract}


\section{INTRODUCTION}
The White Rabbit (WR) technology~\cite{wr_pub_my} and its operation are primarily designed for deployment in optical network infrastructure. Each operational node within the White Rabbit network is typically connected using an SFP (Small Form-factor Pluggable) transceiver, which serves as a converter between the electrical and optical communication domains. This device performs the conversion of electrical signals to optical signals and vice versa.

When dedicated fiber is available for connecting two nodes, either a BiDi (Bidirectional) WDM (wavelength division multiplexing) transceiver or a DWDM (Dense Wavelength Division Multiplexing) transceiver can be used~\cite{tranceiver_book}. Both types of transceivers have their advantages and disadvantages. One of the critical aspects when selecting an appropriate SFP transceiver is its impact on the synchronization accuracy of two White Rabbit nodes.

Specifically, we are interested in the effect of changes in chromatic dispersion caused by varying external environmental conditions on the synchronization accuracy between two White Rabbit nodes, when using either a BiDi WDM transceiver or a DWDM transceiver. The most significant factor influencing the chromatic dispersion of optical fiber is temperature. The temperature of the optical fiber changes its refractive index, which in turn affects the chromatic dispersion~\cite{chromatic_book}.

\section{THEORETICAL BACKGROUND}
BiDi WDM Transceivers use a single optical fiber, where two different wavelengths of the optical signal are operated for communication in each direction. Common wavelength combinations include 1310/1550 nm or 1490/1550 nm. Connecting these transceivers does not impose additional demands on the optical infrastructure. An advantage is that communication occurs over a single optical fiber, ensuring that the total path length the optical signal travels is the same in both directions. However, the significant difference between the transmitting and receiving wavelengths results in differing propagation times due to the mentioned chromatic dispersion.

DWDM Transceivers also use two wavelengths for bidirectional communication. Physically, these two optical signals are partially or entirely routed through two separate optical fibers. If only a single optical fiber is available for connection, there arises a necessity to expand the optical infrastructure with a DWDM Mux/Demux. In such cases, the difference in propagation time in each direction can stem from varying lengths of optical fibers connecting the transceiver to the optical network, different delays for each wavelength on the DWDM Mux/Demux device, and also chromatic dispersion. An advantage of DWDM transceivers is the ability to use wavelengths that are significantly closer together compared to the first technological solution. Consequently, the resulting impact of chromatic dispersion should be considerably smaller. Typically, wavelengths in the range of 1520.25 nm (channel 1) to 1577.03 nm (channel 72) with a channel separation width of 0.8 nm, referred to as the C-band, are used today. For our purposes, two adjacent channels can be used, providing a difference in the transmitting and receiving signal wavelengths of only 0.8 nm, which is significantly better compared to the BiDi WDM technology.

\section{REALIZATION}
\begin{figure}[H]
\centering
\begin{minipage}{0.4\textwidth}
  \centering
  \includegraphics[width=1.0\linewidth]{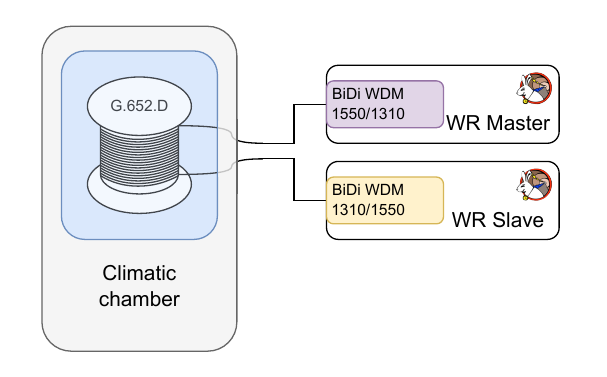}
\end{minipage}%
\hfill\vline\hfill
\begin{minipage}{0.59\textwidth}
  \centering
  \includegraphics[width=1.0\linewidth]{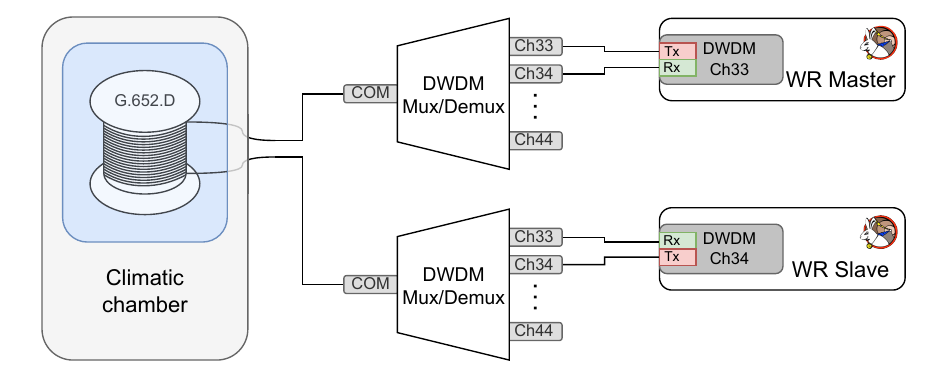}
\end{minipage}
\caption{Optical setup for measurements with BiDi WDM and DWDM transceivers.}
\label{meas:opto_setup}
\end{figure}

A test setup of the optical network was constructed under laboratory conditions to simulate a real optical network. This model used a 20 km long G.652.D optical fiber. The influence of external environmental factors on a real route was simulated using a climatic chamber~\cite{climate_chamber}, where the optical fiber was placed. A simplified diagram of the entire verification setup is shown in Figure \ref{meas:opto_setup}.
For the first measurements, BiDi WDM transceivers operating at wavelengths of 1310/1550 nm were used. For subsequent measurements, DWDM transceivers on channels Ch33 and Ch34, in combination with a DWDM Mux/Demux, were employed.

\begin{figure}[H]
    \centering
    \includegraphics[width=0.75\textwidth]{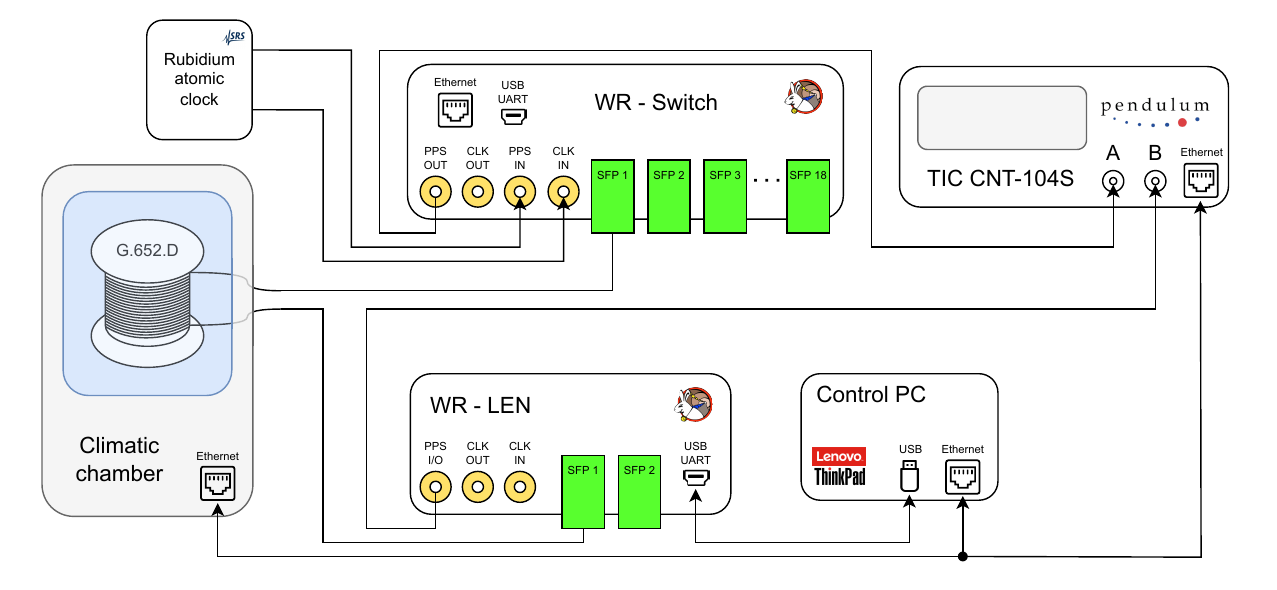}
    \caption{Block diagram of the instrumentation setup for measurement.}
    \label{fig:meas_setup}
\end{figure}

From the perspective of the instrumentation, a WR master in the form of a WRS (White Rabbit Switch)~\cite{wrs_cite} was deployed on one side of the optical fiber, and a WR slave in the form of a WR-LEN device~\cite{wrlen_cite} was used on the other side. For the time reference on the WR master side, a rubidium frequency standard SRS FS725~\cite{rubidium_datasheet} was utilized. The delay between the 1PPS signals of the WR master and WR slave devices is measured using a TIC (time interval counter) Pendulum CNT-104S~\cite{pendulum_cnt_datasheet}. A PC with the necessary software, in the form of Python scripts, is used for controlling the climatic chamber, configuring the WR devices, and storing data obtained from the counter. A block diagram of the instrumentation setup is shown in Figure \ref{fig:meas_setup}.

\section{RESULTS}
\begin{figure}[H]
\centering
\begin{minipage}{0.5\textwidth}
  \centering
  \includegraphics[width=0.95\linewidth]{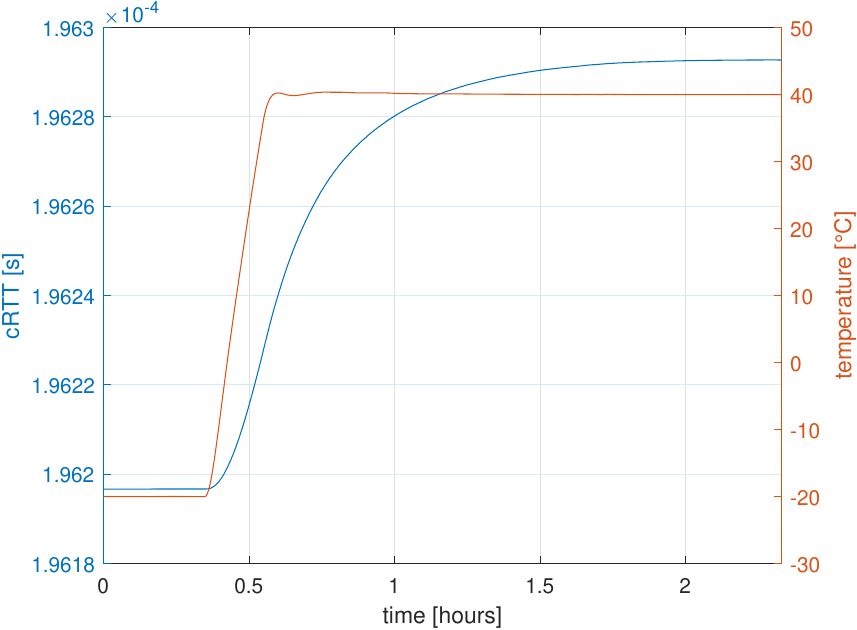}
\end{minipage}%
\begin{minipage}{0.5\textwidth}
  \centering
  \includegraphics[width=0.95\linewidth]{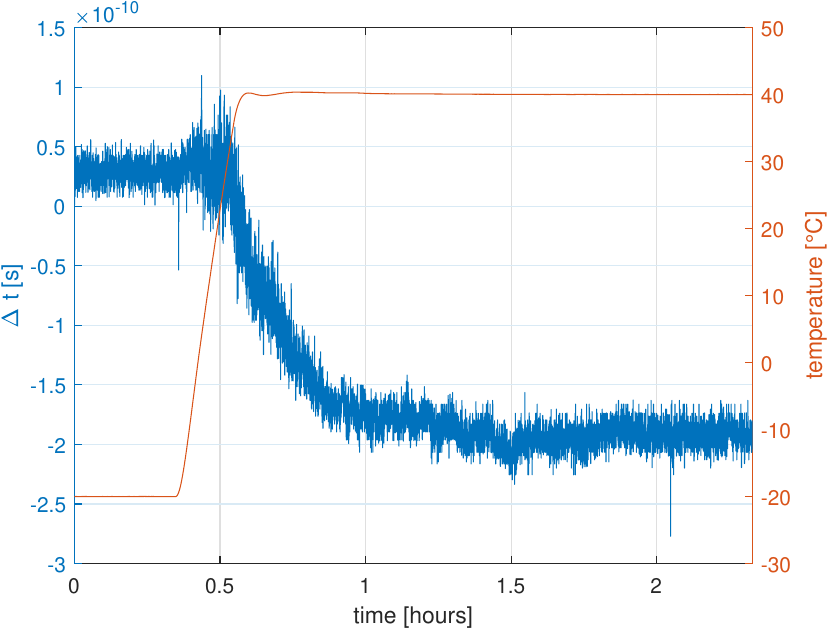}
\end{minipage}
\caption{Measurement results of cRTT and $\Delta t$ versus optical fiber temperature for BiDi WDM transceivers.}
\label{fig:res_dwm}
\end{figure}

\begin{figure}[H]
\centering
\begin{minipage}{0.5\textwidth}
  \centering
  \includegraphics[width=0.95\linewidth]{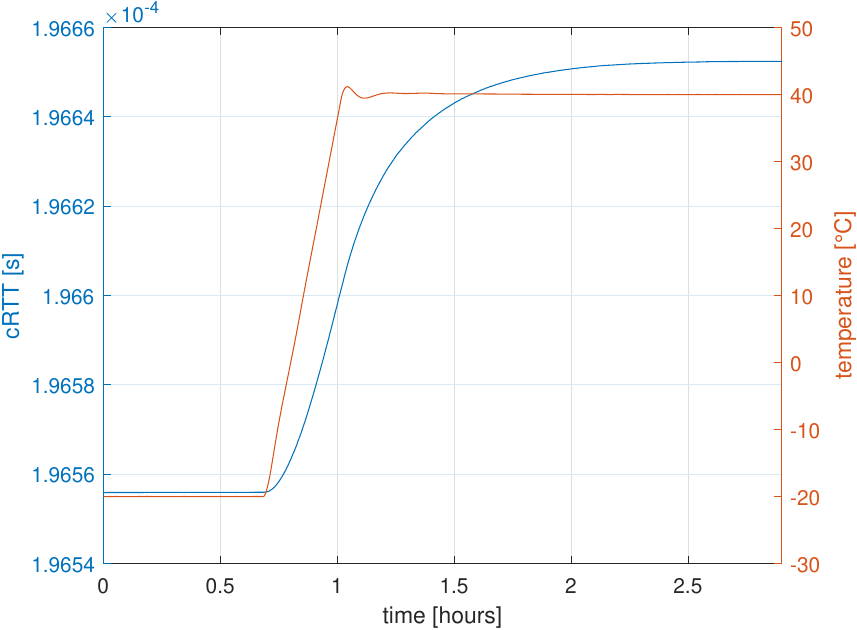}
\end{minipage}%
\begin{minipage}{0.5\textwidth}
  \centering
  \includegraphics[width=0.95\linewidth]{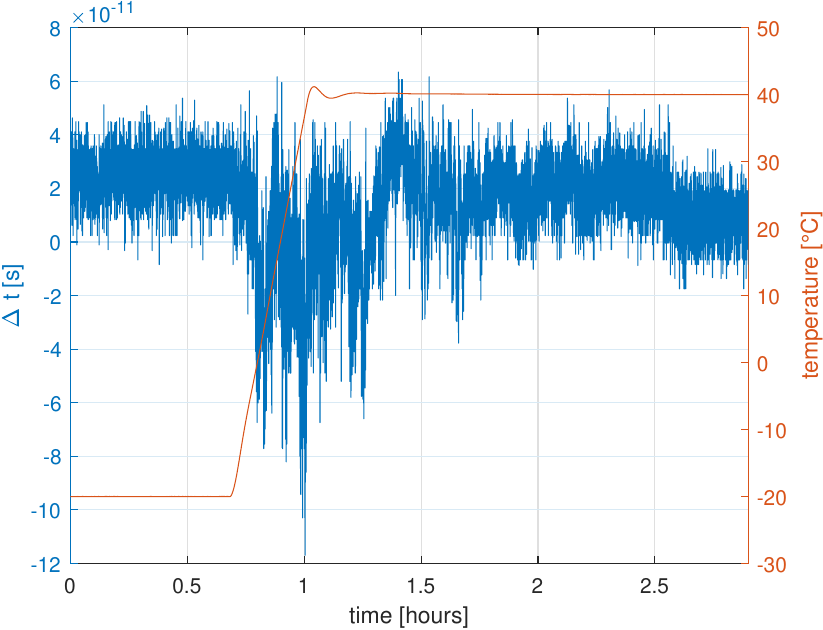}
\end{minipage}
\caption{Measurement results of cRTT and $\Delta t$ versus optical fiber temperature for DWDM transceivers.}
\label{fig:res_dwdm}
\end{figure}

The results of the two conducted experiments are shown in Figures \ref{fig:res_dwm} and \ref{fig:res_dwdm}. For both measurements, the same climatic conditions were maintained, including changes over time. Initially, the climatic chamber was cooled to a temperature of \SI{-20}{\celsius}. After stabilization, the offset of the WR master and slave devices, measured as the time $\Delta t$, was recorded. Following a brief recording period of approximately 30 minutes at this temperature, the climatic chamber was set to heat up to \SI{40}{\celsius}. To assess whether the temperature of the conditioned fiber had stabilized, we simultaneously recorded the value of cRTT (corrected round-trip time). As observed, the temperature of the conditioned fiber stabilized after about 90 minutes. The resulting offset change is evaluated after the fiber temperature has stabilized.
Table \ref{tab:results_values} shows the average offset in the steady state for optical fiber temperatures of \SI{-20}{\celsius} and \SI{40}{\celsius}. The table also includes the final difference in offset between the two WR devices with a \SI{60}{\celsius} change in optical fiber temperature. As can be seen from the processed results, the offset is significantly less affected by temperature when using DWDM transceivers.
\begin{table}[H]
\caption{Final measurement results comparing BiDi WDM and DWDM transceivers technologies.} 
\label{tab:results_values}
\begin{center}       
\begin{tabular}{|l|c|c|c|} 
\hline
\rule[-1ex]{0pt}{3.5ex} Transceivers technology & $\Delta t_{\SI{-20}{\celsius}}~[\SI{}{\pico\second}]$    & $\Delta t_{\SI{40}{\celsius}}~[\SI{}{\pico\second}]$ & $\Delta t_{\Delta\SI{60}{\celsius}}~[\SI{}{\pico\second}]$ \\
\hline
\rule[-1ex]{0pt}{3.5ex}  BiDi WDM               & 29.0                              & -191.4                         & 220.4                        \\
\hline
\rule[-1ex]{0pt}{3.5ex}  DWDM                   & 24.9                              & 7.9                           & 17.0                          \\
\hline
\end{tabular}
\end{center}
\end{table}
\noindent where $\Delta t_{\SI{-20}{\celsius}}$ is the WR master and WR slave time offset with optical fiber temperature at \SI{-20}{\celsius}, $\Delta t_{\SI{40}{\celsius}}$ is the WR master and WR slave time offset with optical fiber temperature at \SI{40}{\celsius}, and $\Delta t_{\Delta \SI{60}{\celsius}}$ is the time offset change due to a \SI{60}{\celsius} optical fiber temperature change.

\section{CONCLUSION}
The aim of this study was to investigate the impact of different transceiver technologies on the offset between two White Rabbit nodes when subjected to changes in external environmental conditions affecting the connecting optical fiber. The transceiver technologies compared were BiDi WDM and DWDM. For the BiDi WDM transceiver, the separation between the Tx and Rx channels was 240 nm, whereas for the DWDM transceiver, the separation was only 0.8 nm.

An experiment involving a 20 km long optical fiber subjected to a \SI{60}{\celsius} temperature change demonstrated that DWDM transceivers offer greater offset stability between the two WR devices. The temperature change of \SI{60}{\celsius} resulted in an offset change of 17.0 \SI{}{\pico\second} for DWDM transceivers and 220.4 \SI{}{\pico\second} for BiDi WDM transceivers. The offset change due to temperature variations in the optical fiber was nearly thirteen times less significant with DWDM transceivers. Therefore, DWDM transceivers should be preferred in WR networks to achieve the best synchronization accuracy.

\section*{ACKNOWLEDGMENTS}
This work was supported by the Grant Agency of the Czech Technical University in Prague, grant number SGS24/062/OHK3/1T/13. The research was also partially supported by project CZ.02.01.01/00/22\_008/0004649 (QUEENTEC).

\bibliographystyle{spiebib} 

\end{document}